# Adaptive mosaic image representation for image processing.

Evgenia Gelman

Ben-Gurion University, Israel

egelman@bgu.ac.il

**Abstract.**

Method for a mosaic image representation (MIR) is proposed for a selective treatment of image fragments of different transition frequency. MIR method is based on piecewise-constant image approximation on a non-uniform orthogonal grid constructed by the following recurrent multigrid algorithm. A sequence of nested uniform grids is built, such that each cell of a current grid is subdivided into four smaller cells for the next grid designing. In each grid the cells are selected, where the color intensity function can be approximated by its average value with a given precision (thereafter 'good' cells). After replacing color of good cells by their approximating constants reconstructed image looks like a mosaic composed of one-colored cells. Multigrid algorithm results in the stratification of the image space into regions of different transition frequency. Sizes of these regions depend on the few tuning precision parameters, that characterizes adaptability of the method to the image fragments of different non-homogeneity degree. The method is found efficient for prominent contour (skeleton) extraction, edge detection as well as for the Lossy Compression of single images and video sequence of images.

1.Inroduction.

The Mosaic Image Reconstruction (MIR) method for image processing introduced in [1] is inspired by the top-efficiency multigrid methods in numerical simulations for fluid dynamics. The ideas of this method are beneficial in widely varying types of problems related to multi-scale representation and computations (see survey [2]). Image processing is a highly suitable field for the multigrid method due to the multi-scale nature of images.

The MIR method is effective for a wide range of the problems aimed to distinguish the image fragments of a different homogeneity degree (transition frequency): prominent contour (skeleton) extraction, edge detection, Lossy Compression, pattern recognition etc. The current study is mainly concerns with the Lossy Compression problem. The application of the method to other problems, handled in much the same way will be also demonstrated. Lossy Compression methods are characterized by the three conflicting requirements on: compression ratio, image quality and coding speed. An efficient approach to obtaining a compromise between these requirements consists in transmitting only information that is essential to human perception. Therefore adaptive methods, i.e. those which select the more informative fragments compared with the less ones, are the most valuable.

Among the most widespread methods of Lossy Compression are: Fractal image compression, Wavelets, JPEG (Joint Photographic Experts Group), ABTC (Adaptive Block Truncation Coding algorithm).

- Fractal method of image representation yields very good results with regard to compression ratio and image quality, but is not so successful in coding speed. Considerable success has been reached in integrating fractal compression with alternative techniques [3].

- The method of wavelets [4] provides useful tools for all the three above mentioned requirements. Both wavelet transforms and MIR method results in splitting image into regions of different homogeneity degree. But algorithms of these two methods are different.

- JPEG is a standardized image compression technique [5]. It is one of the most popular coding techniques in imaging applications ranging from Internet to digital photography. A useful property of JPEG is that the degree of information loss can be varied by adjusting compression parameters, which means that the compression ratio may be sacrificed for the sake of image quality and vice versa.

- The Adaptive Block Truncation Coding algorithm (ABTC) [6, 7] operates by subdividing the image into blocks. Elements of each block are approximated by one or two constants. Due to the grid non-uniformity and the availability of two tuning parameters the method is adaptive to the image fragments.

In the present study a novel Lossy Compression method is proposed and compared with JPEG. The original image space is reconstructed by a piecewise-constant function approximating the color intensity function Φ with a given precision on a non-uniform orthogonal grid constructed by the recurrent multigrid procedure. Reconstructed image looks like a mosaic composed of one-colored cells.

In fact, each compression method is characterized by a compression curve describing the correspondence between compression ratio and image quality. This compression curve is obtained with the help of tuning parameters employed by the user. One turning parameter is used in JPEG, two in the ABTC algorithm and N parameters in MIR method. The availability of the multigrid structure and the tuning precision parameters makes MIR method highly adaptive to the image fragments of different transition frequency Owing to adaptability MIR method is efficient for fast edge detection, prominent contour (skeleton) extraction. In Lossy Compression when MIR method is used, increase in compression ratio along the compression curve does not lead to essential decrease in image quality.

Efficiency of MIR method for video compression is based on mosaic representation of sequential still images and reduction of output information at the expense equality corresponding sells of contiguous images.

In recent years several constructing tools and viewers for panoramic images have appeared as successful methods for video compression [8,9], where panoramic images are represented as mosaic compounded of consequents of still images (in these works notion 'mosaic' should be distinguished from that in MIR method).

The paper is organized as follows. The next Section is devoted to the description of the multigrid procedure, the algorithms for single image compression, video compression and image restoration. In Section 3 ability is discussed of the MIR method for the adaptive control. Section 4 presents examples of implementing the MIR method for single image compression. The results are compared with those obtained by using JPEG. Also applications of the method are presented to compression of the two consecutive video images, edge detection, prominent contour (skeleton) extraction and pattern recognition. Some conclusions are drown in Section 5.

## 2. DESCRIPTION OF THE MIR METHOD.

**2.1 Mosaic representation of a single image.** The color intensity function $\Phi(i, j)$ is given on the image plane $D(i, j)$; i, j are the coordinates of points, i=0, ..., I from top to bottom; j=0, ..., J from left to right.. A sequence of grids $G^n$ is built. The grid $G^n$ consists of $a_n \times a_n$ cells $K^n(p, q)$ with coordinates (p, q). The cell $K^n(p, q)$ consists of the points (i,j) subject to

$$a_n \, p \leq i \leq a_n \, (p+1), \qquad a_n \, q \leq j \leq a_n \, (q+1) \, . \tag{1}$$

Each grid $G^{n+1}$ is nested into preceding one $G^n$ according to the formula

$$a_n = 2a_{n+1}. \qquad (2)$$

Hence each 'parent' cell $K^n$ comprises 4 'children' cells $K^{n+1}$.

The cell $K^n$ (p,q) is termed a $\varepsilon_n$-cell or 'a good cell' if function $\Phi(i, j)$ on it is approximated by the average value $V(p,q)$ with the given precision $\varepsilon_n$.

The algorithm is aimed to find the all good cells for each grid $G^n$. The 'children' of good cells are termed 'processed cells'. The children of the processed cells are also termed processed cells. The remaining cells, i.e. cells that are neither processed nor good are termed 'bad cells' of grid $G^n$. For each grid $G^n$ matrix $M^n$ (p, q) is generated each element of which corresponds to the cell $K^n$ (p,q) and equals to the average value $V^n$ (p,q) for good cells and (-1) for bad cells. In the searching for good cells of the current grid $G^n$ the area of the processed cells of the previous grid has to be ignored. Finally the found values $V^n$ (p,q) will replace function $\Phi(i, j)$ on the good cells. As a result, a new matrix $D(i, j)$ is obtained composed on the cells of different sizes. In each cell one constant approximates all its original elements $\Phi(i, j)$ of matrix $\Phi$.

Notion mosaic implemented in the present study (compare with [8]) implies image constructed of the blocks with the property that they are one-colored, while any bigger region including the block, is multicolored. Due to this property, each one-colored block can be considered as maximal and number of all these blocks minimal. This property of mosaic is useful for the image compression purpose.

Reconstructed by MIR method image is non-uniform grid, where each cell is a one-colored good cell in some grid $G^n$, while its parent is a bad cell in the previous grid $G^{n-1}$ and hence multicolored. Thus, processed by MIR method image is mosaic.

## 2.2. Output information and image restoration

The output information for the entire multigrid procedure are two vectors: $s = (s_k)$ and $f = (f_l)$, the former containing information on the disposition of the good cells within grids $G^n$, the other on the average values $V(p, q)$ on these cells. At every n-th step of the procedure the n-th fragment of these vectors is built. All the fragments are built consequently one after another. The algorithm for building $n$-th fragments is as follows. Scan all cells of grid $G^n$ from left to right and from top to bottom with the counter m=0, 1, ...,,$m^n$, numbering cells $K^n(p_m, q_m)$; where $m^n$ is a number of all cells in the grid $G^n$. Set $s_m = 1$ if $K^n(p_m, q_m)$ is a good cell and $s_m = 0$ otherwise.

Scan cells $K^n(p_l, q_l)$ as described above with the counter $l=0, 1, ..., l^n$, where $l^n$ is a number of good cells, numbering only good cells and setting $f_l = V^n(p_l, q_l)$. For this procedure matrixes $M^n$ are used.

The aim of the procedure for image restoration is to build the function $D(i, j)$ that approximates the original function $\Phi(i, j)$ using the output vectors $s$ and $f$ obtained as a result of multigrid algorithm. The restoration process consists of the same procedure for successive building of the grids $G^1, G^2, ..., G^N$ as in the compression algorithm, with the only difference that in the compression algorithm the information about good cells (their disposition in grid $G^n$ and the average value of elements $\Phi(i, j)$ over them) was transferred to the vectors $s$ and $f$, whereas in restoration algorithm they are transferred in the opposite direction.

### 2.3. Edge and prominent contour (skeleton) detection.

Edges are defined as the regions of the sharpest transitions. They are obtained at the last step of the multigrid algorithm as a collection of all cells for the finest grid.

Skeleton is obtained as a collection of good cells for the few last grids, number of which depends on the application.

### 2.4. Lossy Compression algorithm.

Compression of a single image consists of two steps: mosaic image representation and building of two output vectors $s = (s_k)$ and $f=(f_l)$, providing mosaic image restoration..
Video compression is carried out with mosaic represented still images.
It is based on comparing color intensities of the corresponding cells of the same level grids for two consecutive images. If their values are equal, then color intensity in the second image is replaced by zero.
For restoration purposes, zero value of the color intensity in the second image is replaced by the value of the corresponding cell in the first image.

### 2.5 Pattern recognition based on skeletons comparing.

If it is required to locate a given model (pattern) inside a definite picture, the follow algorithm is proposed:

1) mosaic representation both the model and the picture;

2) construction of their skeletons;

3) moving model over the picture counting number of coincident points. These numbers are the values of the function depending of the model's locations. Indicator of right location is maximal value for gradient of the function.

## 3. ADAPTIVE CONTROL IN THE MIR METHOD.

The compression ratio and the image quality depend on the precision parameters $\varepsilon_n$. The greater $\varepsilon_n$, the wider is an area covered by good cells and consequently the higher the compression ratio. But this worsens the image quality. So, the control factor for $\varepsilon_n$ is the image quality of the processed area. The precision parameters are either user-defined or chosen automatically according to some other criteria.

If a region is covered by the uniform cells, then the size of the cells determines texture of the region. The region of a definite texture is constructed in each grid $G^n$ as a space covered by good cells. The proposed multigrid algorithm automatically stratifies the image space into regions of different textures. Areas with the lower transition frequency are processed in the grids of the bigger cells than areas with the higher transition frequency. Hence MIR method splitting the image into regions of different textures provides hereby separation the image space into regions of different transition frequency. We consider this property as the adaptability of the method to the image fragments of different transition frequency. This property is enhanced by the possibility to change $\varepsilon_n$ from step to step of the multigrid algorithm.

For example, if it is required to pick out details, which are most bright for the human perception, we can gain it isolating regions of the sharpest transitions by increasing parameter $\varepsilon_n$. Then desired details will be revealed in the last steps of multigrid algorithm. Thus, in the Lossy Compression problem the parameters $\varepsilon_n$ act as controlling parameters, the tuning of which enables the desirable compromise between the compression ratio and the image quality. Areas with the sharpest transitions, are processed at the last step.

## 4. APPLICATIONS.

**Figure 1.** The examples demonstrate a comparison MIR and JPEG methods in Lossy Compression problem. Fig. 1 shows image processed by MIR method with the parameters

$\varepsilon_n = 30$ (Image 30.bmp) and $\varepsilon_n = 50$ (Image 30.bmp) at n=16, 8, 4, 2, 1.

Each of the mosaics is matched with the results of the image processing by the JPEG method yielding the same compression ratio.

Image 30.bmp and Image 91.jpeg have compression ratio 39

Image 50.bmp and Image 99.jpeg have compression ratio 58

Numbers 91 and 99 in the names of the images are tuning parameters of JPEG method. It can be seen that at high compression ratios the image processed by MIR method is superior in image quality, while not being inferior at the lower compression ratios.

**Figure 2** shows the use of the method as edge detector. Two values $\varepsilon_0$ and $\varepsilon_1$ were set for this purpose: $\varepsilon_n = \varepsilon_0$ for n=16, 8, 4 and $\varepsilon_n = \varepsilon_1 < \varepsilon_0$ for n=2. Only the area processed for n=2 and n=1 is marked by original color.

**Figure 3** shows prominent contour constructed on the two last steps of multigrid algorithm.

**Figure 4** shows two consecutive video images **a** and **b**. The annexed pictures show their mosaic representations beside images **a** and **b** correspondently. The image in the lower right corner is the difference between the two mosaic images. It is used instead of the image **b** for compression.

**Figure 5.**

**Problem .** A set of pictures is given together with one small picture (model), which presents a fragment taken from all the bigger pictures, but somewhat modified in color and shape. It is required to locate the model inside each frame. Sometimes several models are given for comparison and it is required to find the one that is inside a definite frame. The program is to operate in real time. The algorithm consists in comparing the skeletons of the pictures and the model. The annexed pictures show the original frame beside the same frame with the model identified inside it by white borders. Three models were given for the search.

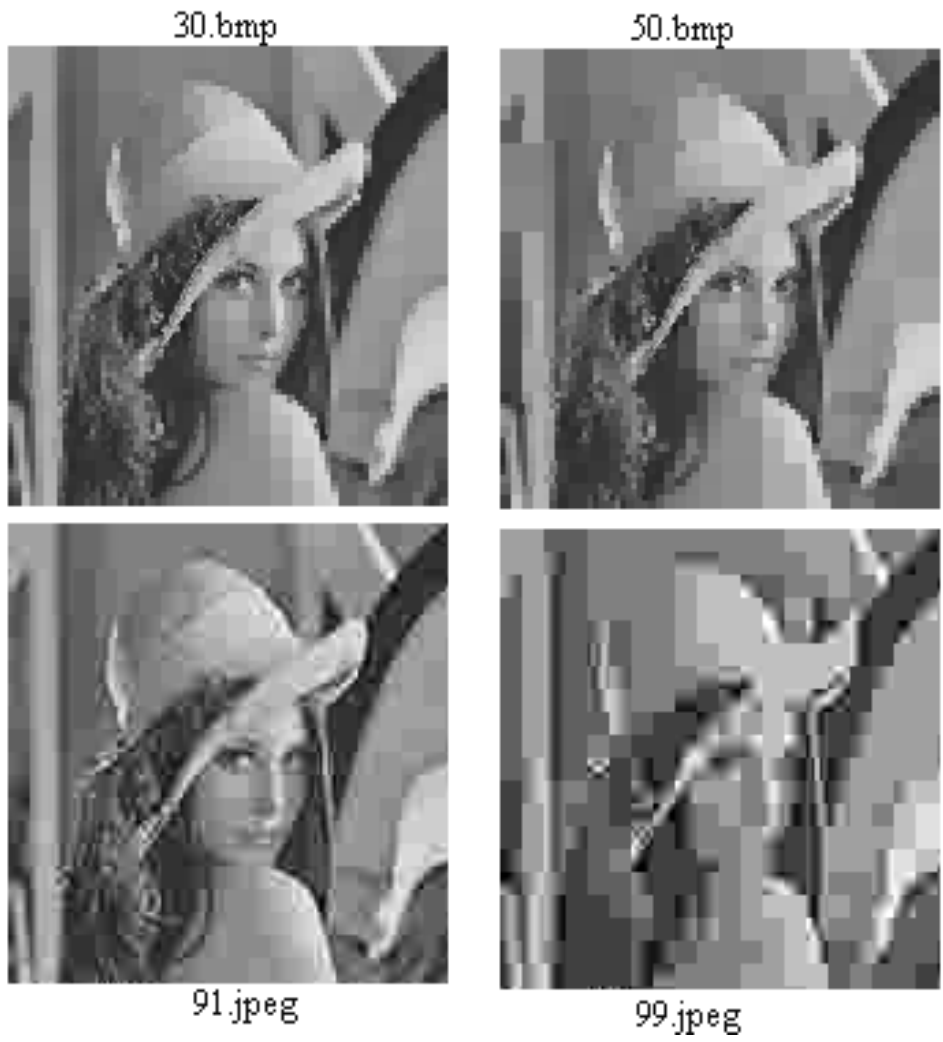

Figure 1.

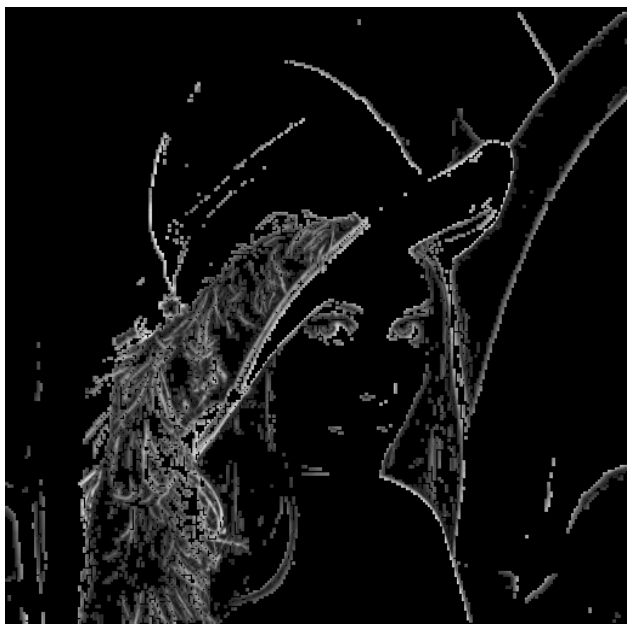

Figure 2

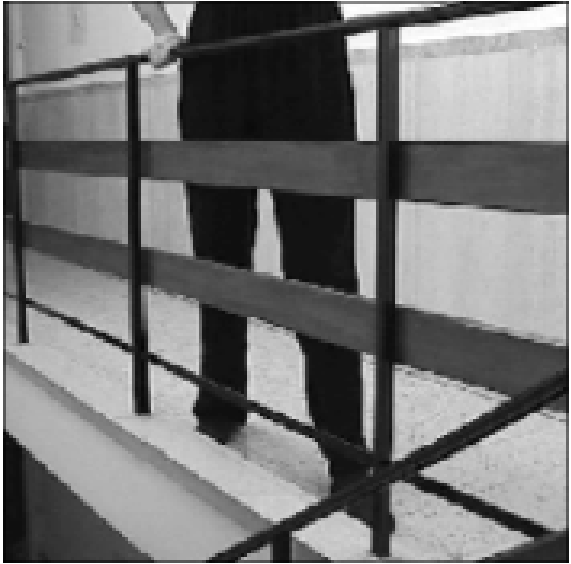
Figure 3a . Original image.

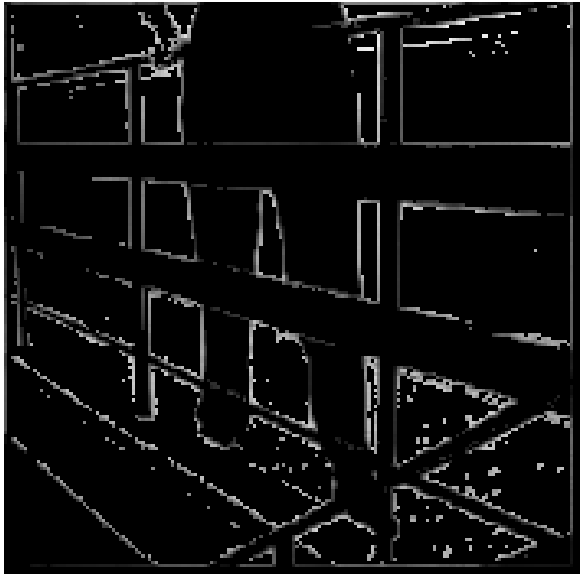

Figure 3b. Prominent contour.

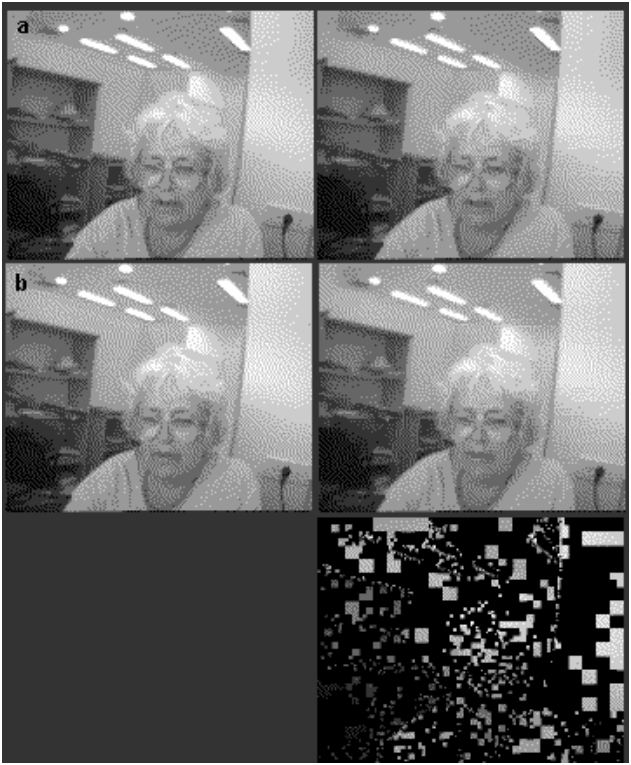

Figure 4.

Figure 5.

**Models**

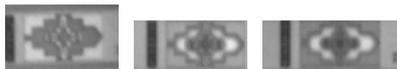

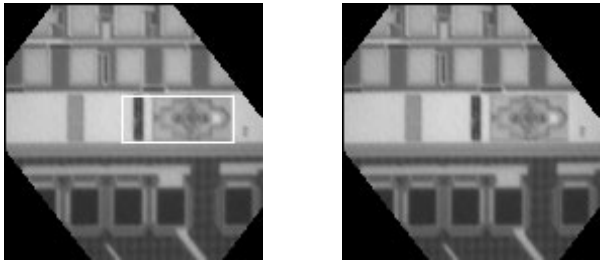

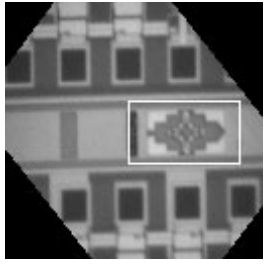
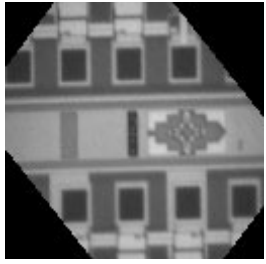

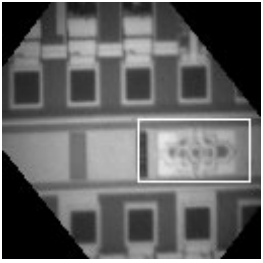
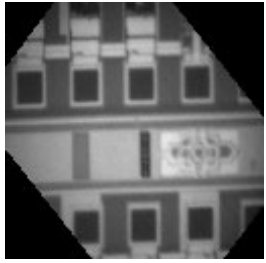

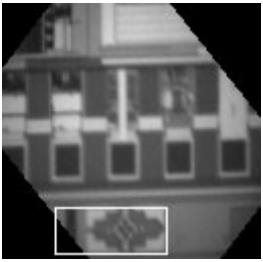
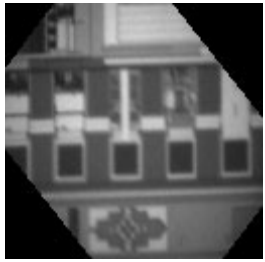

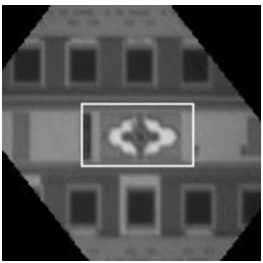
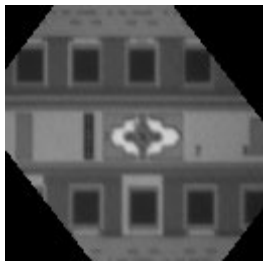

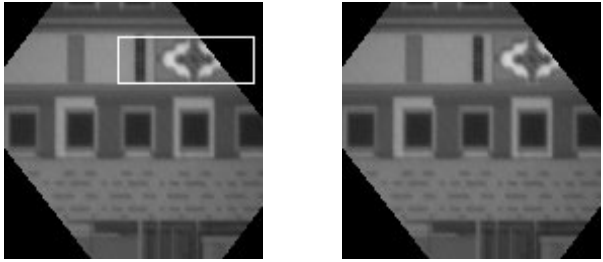

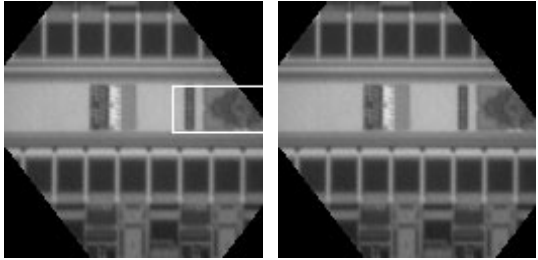

## 5. CONCLUSIONS

A general novel approach to the range of image processing problems is proposed. The method is based on the approximation of the color intensity function by a piecewise-constant function with the help of a special multigrid technique. It produces a mosaic post-processed image. The image space is stratified into areas with smooth and sharp transitions of color intensity, thereby enabling selection areas that are most essential for human perception. Availability of several tuning parameters and their selective treatment for different image fragments provides a high level of adaptability of the method. This property enables the method to be used for edge detection, prominent contours extraction and Lossy Compression for individual images and video consequence images. Numerical simulations have shown that MIR method excels JPEG in image quality when high compression ratio is required and is on about the same level at low compression ratio. The preferable images for using MIR method for Lossy Compression are those which contain areas with smooth and sharp transitions of color intensity.

**Acknowledgement.**

I am grateful to Dr. Yuri Shtemler for his superb assistant and encouragement throughout this work.